\documentclass[%
 aip,
 amsmath,amssymb,
 reprint,%
]{revtex4-1}

\usepackage{graphicx}% Include figure files
\usepackage{dcolumn}% Align table columns on decimal point
\usepackage{bm}% bold math
\usepackage[normalem]{ulem}

\usepackage[utf8]{inputenc}
\usepackage[T1]{fontenc}
\usepackage{mathptmx}
\usepackage{etoolbox}
\usepackage{xcolor}
\usepackage[colorlinks=true, linkcolor=blue, citecolor=blue, urlcolor=blue]{hyperref}
\usepackage{physics}

\newcommand{\beginsupplement}{%
  \clearpage
  \onecolumngrid
  \setcounter{section}{0}
  \setcounter{subsection}{0}
  \setcounter{figure}{0}
  \setcounter{table}{0}
  \setcounter{equation}{0}
  \renewcommand{\thesection}{S\arabic{section}}
  \renewcommand{\thesubsection}{S\arabic{section}.\arabic{subsection}}
  \renewcommand{\thefigure}{S\arabic{figure}}
  \renewcommand{\thetable}{S\arabic{table}}
  \renewcommand{\theequation}{S\arabic{equation}}
}

\makeatletter
\def\@email#1#2{%
  \patchcmd{\titleblock@produce}
    {\frontmatter@RRAPformat}
    {\frontmatter@RRAPformat{\produce@RRAP{*#1: \href{mailto:#2}{#2}}}\frontmatter@RRAPformat}
    {}{}
}%
\makeatother

\begin{document}

\title{Impact of dynamic electrostatic disorder on hole mobility in rubrene: a nonadiabatic molecular dynamics investigation}

\author{Jan Elsner}
\affiliation{Department of Physics and Astronomy and Thomas Young Centre, University College London, London WC1E 6BT, U.K.}
\affiliation{Lehrstuhl f\"ur Theoretische Chemie II, Ruhr-Universit\"at Bochum, 44780 Bochum, Germany}
\affiliation{Research Center Chemical Sciences and Sustainability, Research Alliance Ruhr, 44780 Bochum, Germany}
\author{Samuele Giannini}
\affiliation{Department of Chemistry and Industrial Chemistry, University of Pisa, 56124 Pisa, Italy}
\author{Jochen Blumberger}
\affiliation{Department of Physics and Astronomy and Thomas Young Centre, University College London, London WC1E 6BT, U.K.}

\makeatletter
\@email{Corresponding author}{j.blumberger@ucl.ac.uk}
\makeatother

\date{\today}

\begin{abstract}
High-mobility organic molecular crystals such as rubrene are important materials for organic electronics, yet a quantitatively predictive description of their charge transport properties remains challenging.
Direct mixed quantum-classical nonadiabatic molecular dynamics simulations provide a promising route by explicitly propagating the charge carrier wavefunction, without assuming a specific transport mechanism. However, previous large-scale simulations of apolar molecular crystals have commonly neglected dynamic electrostatic disorder, since evaluation of electrostatic interactions is computationally demanding and the approximation appears plausible for apolar systems. 
Here, we use the damped shifted-force (DSF) real-space electrostatic summation method, combined with an efficient addition-subtraction scheme, to include dynamic electrostatic disorder in fragment orbital-based surface hopping (FOB-SH) simulations of room-temperature hole transport in rubrene. 
We find that electrostatic interactions increase the reorganization energy for (hypothetical) nearest neighbour hopping by 29 and 39~meV along the a and b-directions, respectively, relative to a baseline value of 152~meV obtained without electrostatics.
In FOB-SH simulations, electrostatic interactions lead to increased site energy disorder, reducing  the spatial extent  of the hole wavefunction, as measured by a decrease in the inverse participation ratio from 13 to 9, and lowering the predicted mobility along the high-mobility direction from $35$ to $21~\mathrm{cm^2 V^{-1}s^{-1}}$, in close agreement with experiment. 
\end{abstract}

\maketitle

\section{Introduction}

Organic semiconductors have attracted sustained interest as active materials for flexible electronics, thin-film transistors, light-emitting diodes, and photovoltaic devices\cite{myers2012organic, ling2018organic, xie2025high}. Their performance in these applications is often limited by the mobility of charge carriers, making the development of predictive theoretical approaches for charge transport a central goal in the field. Rubrene is a particularly important benchmark system in this context: high-quality rubrene single crystals exhibit among the highest hole mobilities reported for molecular organic semiconductors\cite{podzorov2004intrinsic, hulea2006tunable, xie2013high}.
Rubrene has also been used as a platform to explore spin-transport\cite{carey2025long, Wang2026} and triplet-upconversion\cite{Izawa2021, Sakamoto2022}.

Charge transport in high-mobility molecular crystals such as rubrene occurs in an intermediate regime between the limiting cases of small-polaron hopping and band transport\cite{ciuchi2011transient, fratini2016transient, fratini2020charge}. In this regime, the charge carrier is neither fully localized on a single molecule nor completely delocalized over the crystal. Instead, the picture that has emerged over recent years is one of
transient delocalization\cite{giannini2020flickering,giannini2023transiently}, in which the charge carrier wavefunction forms a dynamic polaron delocalized over several molecules, with its spatial extent continuously modulated by molecular and lattice vibrations and transitions between the densely spaced valence or conduction band states. Here, it is important to note that the (partial) localization is not caused by a change in bond length in response to the presence of a charge carrier, as is the case for small polaron formation in oxides\cite{mckenna2012crossover}, but due to the strong thermal fluctuations of electronic coupling and on-site energy in organic materials (i.e. off-diagonal and diagonal electron-phonon coupling). 
This underlies the characteristic band-like decrease of mobility with temperature in ultrapure high mobility organic crystals\cite{giannini2023transiently}, in contrast to the activated transport commonly observed for small polarons in oxides.

Accurately describing this regime requires a method that does not impose either a fully localized hopping picture or a fully delocalized band picture. Approaches based on kinetic rate theories, such as kinetic Monte Carlo (KMC) simulations using Marcus or Marcus--Levich--Jortner rates\cite{Vehoff2010,Jiang2016a}, are computationally efficient, but assume that charge carriers are localized and move through a sequence of incoherent hopping events. They are therefore ill-suited to describe high-mobility molecular crystals such as rubrene, where carrier delocalization plays an important role.
A second class of methods explicitly treats the coupled electronic and nuclear dynamics, including Ehrenfest dynamics,\cite{Troisi2011c,troisi2007prediction} time-dependent density matrix renormalization group (TD-DMRG),\cite{Li2020} quantum Monte Carlo techniques,\cite{DeFilippis2015} and, more recently, advanced surface-hopping schemes such as MASH.\cite{Runeson2024} Although these approaches provide a more faithful description of charge transport, their computational cost has generally restricted the most accurate quantum-dynamical treatments to one-dimensional or otherwise simplified model systems. Furthermore, many existing studies lack a fully atomistic description of the vibrational degrees of freedom, with environmental and electrostatic effects often incorporated only implicitly through model parametrization. These considerations motivate the use of methods that combine nonadiabatic dynamics with an efficient atomistic description of molecular motion in extended crystal models.

The fragment orbital-based surface hopping (FOB-SH) method is a promising approach to this end\cite{spencer2016fob}. 
FOB-SH is a fully atomistic method in which
an excess charge carrier is propagated quantum mechanically in a basis of molecular fragment orbitals, while the nuclei evolve classically according to fewest-switches surface hopping\cite{tully1990molecular}. The resulting time-dependent electronic Hamiltonian contains diagonal site energies and off-diagonal electronic couplings, both of which fluctuate due to thermal nuclear motion. This allows FOB-SH to capture the competition between electronic coupling, which promotes spatial delocalization of the charge carrier, and energetic disorder arising from local and nonlocal electron-phonon coupling, which tends to localize the charge carrier.

In previous work\cite{elsner2024thermoelectric}, FOB-SH simulations were used to compute hole transport in rubrene using a molecular model in which atomic partial charges were set to zero, thereby neglecting Coulombic electrostatic interactions. 
This approximation is a plausible starting point for apolar molecular crystals such as rubrene, while also being computationally attractive because conventional Ewald-type evaluation of Coulombic interactions becomes a significant bottleneck for the large system sizes required in FOB-SH simulations.
However, neglecting electrostatic interactions removes electrostatic contributions to the reorganization energy and suppresses electrostatic fluctuations of molecular site energies.
Since such site energy fluctuations promote dynamic localization of the charge carrier, it remains unclear to what extent their omission affects carrier delocalization and charge mobility.

Here, we address this question by incorporating electrostatic interactions into FOB-SH simulations of hole transport in rubrene. 
We employ the damped shifted-force (DSF) real-space electrostatic summation method\cite{fennell2006ewald}, using a recently developed addition-subtraction scheme that enables efficient evaluation of the electrostatic contributions to the site energies and forces required in FOB-SH simulations\cite{giannini2025efficient}.
This approach avoids the prohibitive cost of conventional Ewald summation when site energies must be evaluated for many possible charge-localized states, making large-scale nonadiabatic simulations with electrostatic interactions feasible.

We first quantify how Coulombic electrostatics affects the reorganization energy for hole transfer in rubrene, using both the DSF method and smooth particle-mesh Ewald\cite{essmann1995smooth} (SPME) calculations as a reference. 
We show that electrostatic interactions moderately increase the reorganization energy of rubrene (by about 20\%) by enhancing fluctuations of the vertical energy gap between charge-localized diabatic states.
Spectral analysis of the vertical energy gap fluctuations reveals contributions from both low-frequency lattice motions and higher-frequency intramolecular modes. 
We then perform FOB-SH simulations including DSF electrostatics and compare the resulting transport properties with earlier simulations without electrostatics\cite{elsner2024thermoelectric}. Inclusion of electrostatic interactions increases site energy disorder, reduces the spatial delocalization of the hole wavefunction, and lowers the predicted room-temperature hole mobility, bringing it into close agreement with experimental results of Podzorov \emph{et al.}\cite{podzorov2004intrinsic}. These results demonstrate that electrostatic site energy disorder is essential for quantitatively predictive simulations of charge transport in 
molecular organic semiconductors, even if they are composed of apolar molecules such as rubrene.

\section{Results}

\subsection{Reorganization Energy}
\label{sec:results_reorganization_energy}

We first investigate the effect of electrostatic interactions on the reorganization energy for hole transfer in rubrene, a central quantity controlling charge carrier dynamics. 
We emphasize that hole transfer from one rubrene molecule to another
is a fictitious process, since the hole is partially delocalized over several molecules in this material, but one that is important to investigate theoretically because it defines the reorganization energy (Eqs.~\ref{eq:lambda_st} and~\ref{eq:lambda_var}).

The reorganization energy measures the energetic cost associated with nuclear relaxation upon charge transfer and reflects the strength of local electron-phonon coupling.
It contains an internal contribution from structural relaxation of the molecule carrying the charge, as well as an external contribution from the response of the surrounding crystalline environment. 
Coulombic interactions are therefore expected to affect the reorganization energy, in particular its external contribution, by modifying the electrostatic response of the crystal environment to changes in charge localization.

\begin{figure}[t]
\centering
  \includegraphics[width=0.8\columnwidth]{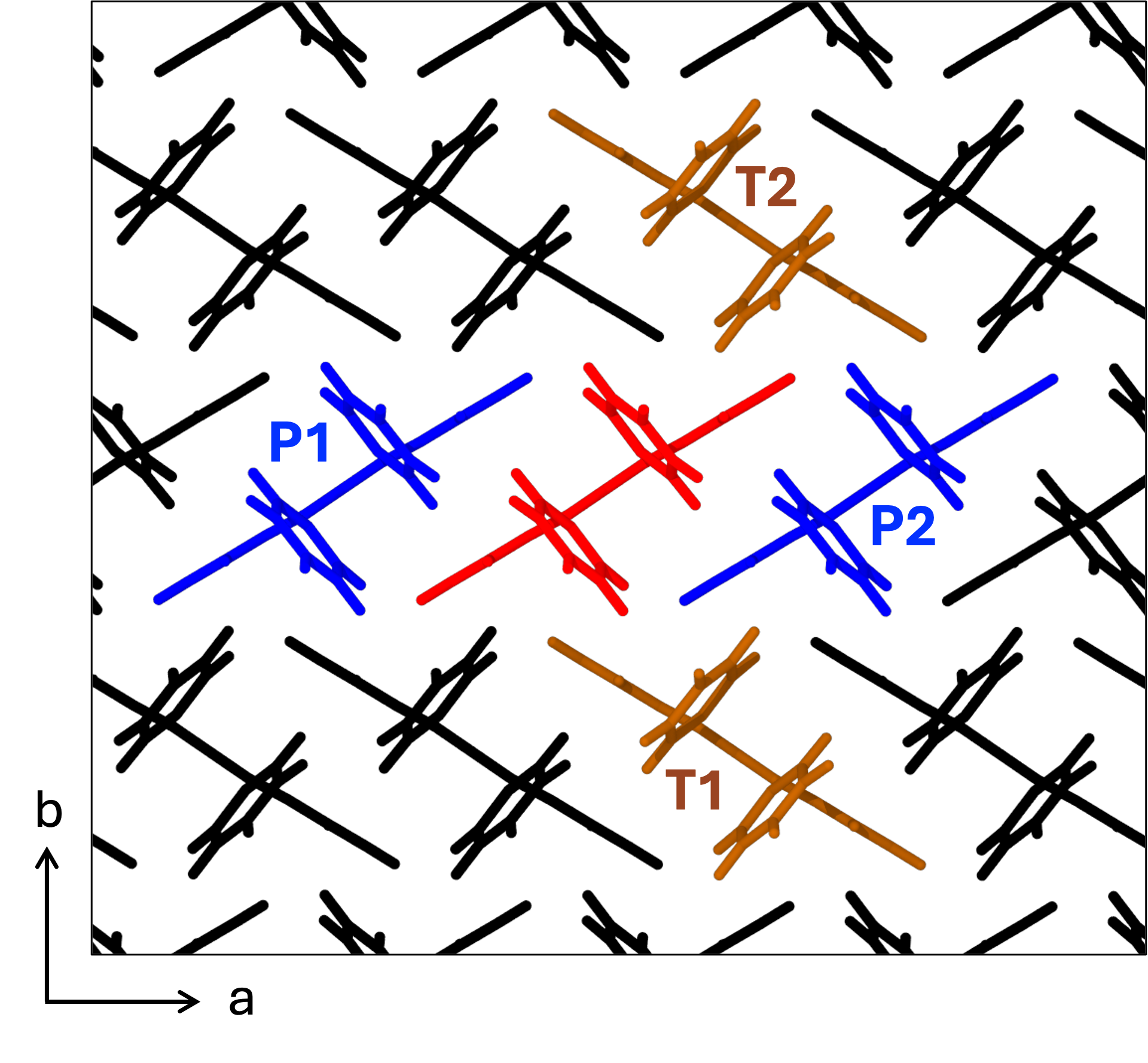}
  \caption{Top-down view of the high-mobility 
  \((a, b)\) plane of crystalline rubrene, showing the molecular pairs used for the reorganization energy calculations. The central molecule shown in red is the molecule on which the hole is localized in the reference diabatic state. The neighbouring molecules used to define vertical energy gaps are highlighted in blue for the \(P\)-type dimers (\(P1\), \(P2\)) and brown for the \(T\)-type dimers (\(T1\), \(T2\)). The \(P\) and \(T\) dimers are stacked along the high- and low-mobility transfer directions, respectively.
}
  \label{fgr:crystal_P_T}
\end{figure}

To quantify this effect, the reorganization energy for hole transfer from molecule \(M\) to neighbour \(N\) in rubrene was calculated using three treatments of Coulombic interactions: a zero-charge model, in which all atomic partial charges were set to zero; the DSF method; and the SPME method.
In each case, reorganization energy was computed according to Eq.~\ref{eq:lambda_st} and Eq.~\ref{eq:lambda_var} from the distribution of vertical energy gaps \(\Delta E_{M\rightarrow N}\) (Eq.~\ref{eq:vertical_energy_gap}) over a molecular dynamics trajectory in the diabatic state corresponding to the hole being localized on molecule \(M\) (see Section~\ref{sec:methods_reorganization_energy} for details). 
Furthermore, reorganization energies were computed for hole transfer from a reference molecule \(M\) to neighbouring molecules along two distinct transfer directions, denoted \(P\) and \(T\). For each direction, two symmetry-related neighbours of \(M\) were considered: \(P1\) and \(P2\) for the \(P\) direction, and \(T1\) and \(T2\) for the \(T\) direction. These molecules are illustrated in Figure~\ref{fgr:crystal_P_T}, where molecule \(M\) is coloured in red, \(P1/P2\) in blue, and \(T1/T2\) in brown. 

\begin{table}[htbp!]
\centering
\caption{
Computed reorganization energies for hole transfer in rubrene using different treatments of electrostatic interactions; a zero-charge model (i.e., no Coulombic interactions), the DSF method and the SPME method. 
}
\begin{tabular}{llcc}
\hline
Method & Dimer type$^a$ &
$\lambda^{\mathrm{st}}$ / meV$^{b}$ &
$\lambda^{\mathrm{var}}$ / meV$^{c}$ \\
\hline
Zero charge & \(P\) & $152$ & $155 \pm 2$ \\
Zero charge & \(T\) & $152$ & $155 \pm 5$ \\
DSF & \(P\) & $175$ & $174 \pm 3$ \\
DSF & \(T\) & $182$ & $187 \pm 4$ \\
SPME & \(P\) & $181$ & $190 \pm 3$ \\
SPME & \(T\) & $191$ & $197 \pm 3$ \\
\hline
\end{tabular}
\begin{flushleft}
$^{a}$ \(P\) and \(T\) denote the neighbour directions considered for hole transfer from molecule $M$. 
Reported values are averages over two symmetry-related neighbouring molecules, see Fig.~\ref{fgr:crystal_P_T}. Values for the individual dimers are reported in Supplementary Section~1. \\
$^{b}$ Stokes shift reorganization energy, $\lambda^\mathrm{st}$, Eq.~\ref{eq:lambda_st}. All statistical uncertainties for $\lambda^\mathrm{st}$ are smaller than 1~meV.\\
$^{c}$ Fluctuation-derived reorganization energy, $\lambda^\mathrm{var}$, Eq.~\ref{eq:lambda_var}, and statistical uncertainty (see~Section~\ref{sec:methods_reorganization_energy} and Supplementary Section~1).
\end{flushleft}
\label{tab:lambda_reorganization_summary}
\end{table}

The computed reorganization energies $\lambda^{\mathrm{st}}$ and $\lambda^{\mathrm{var}}$ for the different treatments of Coulombic interactions are listed in Table~\ref{tab:lambda_reorganization_summary}. Reported \(P\)- and \(T\)-direction values for the reorganization energy are averages over the two corresponding dimers, while values for the individual dimers are given in Supplementary Section~1.  
In the zero-charge model, the reorganization energy is isotropic with respect to hole transfer direction, with \(\lambda^{\mathrm{st}} = 152\)~meV and \(\lambda^{\mathrm{var}} = 155\)~meV for both \(P\) and \(T\).
This is expected because, without electrostatics, the reorganization energy is dominated by intramolecular reorganization, which is independent of the hole transfer direction.
Including electrostatics moderately increases the reorganization energy, with both DSF and SPME giving values larger than the zero-charge model by approximately 20-40~meV, depending on the transfer direction and treatment of electrostatic interactions.

The two estimates, \(\lambda^{\mathrm{st}}\) and \(\lambda^{\mathrm{var}}\), remain within approximately 5\% of one another in all cases, indicating that the vertical energy gap fluctuations are reasonably well described by linear response\cite{blumberger2015recent} (in the linear response limit, \(\lambda^{\mathrm{st}} = \lambda^{\mathrm{var}}\)). 
The DSF values are systematically lower than the corresponding SPME values, but recover most of the electrostatic increase relative to the zero-charge model. Finally, including Coulombic interactions introduces a small directional anisotropy that is absent in the zero-charge model, with the \(T\) direction showing a slightly larger reorganization energy than the \(P\) direction for both electrostatic methods.

\begin{figure*}[t]
\centering
\includegraphics[width=0.8\textwidth]{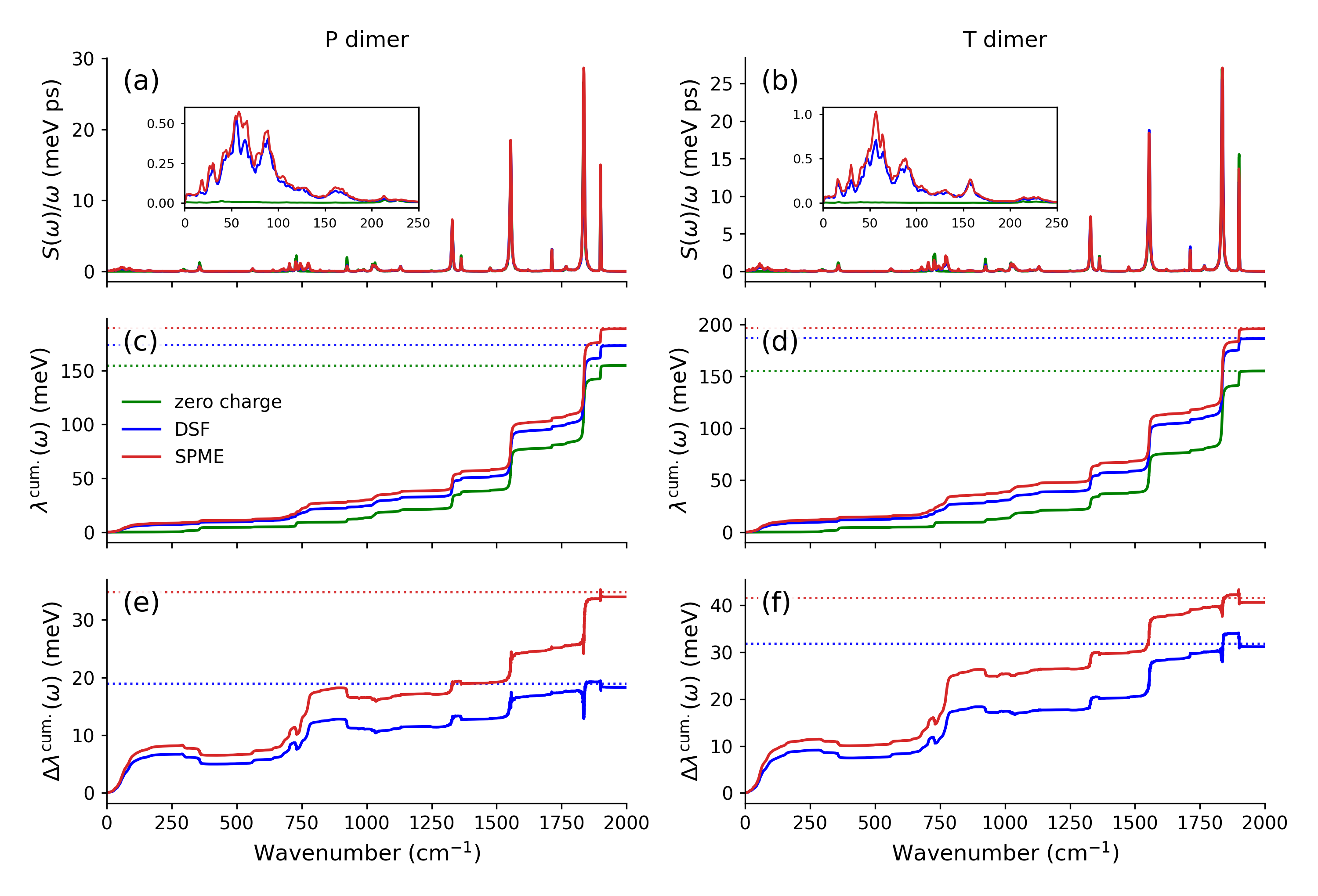}
  \caption{
Spectral density functions of vertical energy gap fluctuations and cumulative reorganization energies for hole transfer, comparing different treatments of Coulombic interactions; zero-charge model in green, the DSF method in blue and the SPME method in red. 
(a,b) Spectral densities, computed using Eq.~\ref{eq:spectral_density}, for transfer along the \(P\) and \(T\) directions, respectively. Insets highlight the low-frequency region from 0 to 250~cm\(^{-1}\).
(c,d) Cumulative fluctuation-derived reorganization energies up to frequency $\omega$, \(\lambda^\mathrm{cum}(\omega)\), computed using Eq.~\ref{eq:lambda_cumulative}, for transfer along the \(P\) and \(T\) directions, respectively. The dotted horizontal lines indicate the corresponding $\lambda^\mathrm{var}$ values obtained from the full energy-gap distribution.
(e,f) Electrostatic contributions to the cumulative reorganization energy $\Delta\lambda^{\mathrm{cum}}(\omega) = \lambda^{\mathrm{cum}}_{j}(\omega) - \lambda^{\mathrm{cum}}_{\mathrm{zero\ charge}}(\omega)$ ($j = \text{DSF, SPME}$ in blue and red, respectively), for transfer along the \(P\) and \(T\) directions, respectively. Dotted lines indicate the values of $\Delta\lambda^{\mathrm{var}} = \lambda^{\mathrm{var}}_{j} - \lambda^{\mathrm{var}}_{\mathrm{zero\ charge}}$ ($j = \text{DSF, SPME}$ in blue and red, respectively).
}
  \label{fgr:spectral_density_distributions}
\end{figure*}

To resolve how different vibrational frequencies contribute to the fluctuation-derived reorganization energy, we computed the spectral density of the vertical energy gap fluctuations, \(S(\omega)/\omega\) (Eq.~\ref{eq:spectral_density}; see Section~\ref{sec:methods_reorganization_energy}). Figures~\ref{fgr:spectral_density_distributions}(a) and (b) show \(S(\omega)/\omega\) for the \(P\) and \(T\) transfer directions, respectively, for each treatment of Coulombic interactions. The peak positions are very similar for the \(P\) and \(T\) directions, indicating that the same vibrational modes dominate the energy-gap fluctuations in both cases.

The insets highlight the low-frequency region below $200\text{ cm}^{-1}$. In the zero-charge model, the spectral density in this region is negligible, whereas both DSF (blue) and SPME (red) show a pronounced low-frequency contribution and are in close agreement. 
This low-frequency enhancement indicates that Coulombic interactions increase the coupling between the localized charge and low-frequency nuclear motions of the rubrene crystal.

To quantify how spectral contributions accumulate into the total fluctuation-derived reorganization energy (Eq.~\ref{eq:lambda_var}), Figures~\ref{fgr:spectral_density_distributions}(c) and (d) show the cumulative reorganization energy, \(\lambda^{\mathrm{cum}}(\omega)\), for the \(P\) and \(T\) transfer directions, respectively. As defined in Eq.~\ref{eq:lambda_cumulative}, \(\lambda^{\mathrm{cum}}(\omega)\) is obtained by integrating \(S(\omega)/\omega\) up to frequency \(\omega\), and therefore gives the portion of \(\lambda^{\mathrm{var}}\) arising from vertical energy gap fluctuations including all modes up to $\omega$.

In Figures~\ref{fgr:spectral_density_distributions}(c) and (d), \(\lambda^{\mathrm{cum}}(\omega)\) increases through a series of step-like changes that coincide with peaks in the spectral density (panels (a) and (b)). 
The DSF and SPME curves follow the same overall profile as the zero-charge model, indicating that most of the dominant vibrational features responsible for site energy fluctuations are present without Coulombic interactions. 
However, several peaks are either absent or significantly weaker in the zero-charge model, leading to a slower accumulation of \(\lambda^{\mathrm{cum}}(\omega)\) compared with the electrostatic simulations, for instance in the low-frequency range \(0-200~\mathrm{cm}^{-1}\) as discussed above (see insets of panels (a) and (b)).

The comparison with the zero-charge model can be seen more directly in Figures~\ref{fgr:spectral_density_distributions}(e) and (f), which plot the excess cumulative reorganization energy, $\Delta\lambda^{\mathrm{cum}}(\omega) = \lambda^{\mathrm{cum}}_{j}(\omega) - \lambda^{\mathrm{cum}}_{\mathrm{zero\ charge}}(\omega)$ ($j = \text{DSF, SPME}$. 
These difference curves show that the electrostatic enhancement of \(\lambda^{\mathrm{var}}\) is distributed over several distinct frequency ranges, with prominent contributions below \(200~\mathrm{cm}^{-1}\), around \(700-750~\mathrm{cm}^{-1}\), and near \(1550~\mathrm{cm}^{-1}\). For the \(P\) direction, the SPME calculation shows an additional increase 
at higher frequencies.

The difference plots also clarify where the DSF and SPME treatments begin to diverge. For both transfer directions, the first noticeable separation appears around \(700-750~\mathrm{cm}^{-1}\). In the \(T\) direction, this discrepancy remains approximately constant over higher frequencies, whereas in the \(P\) direction it increases further 
at higher frequencies.
Thus, DSF captures most of the electrostatic contribution to the cumulative reorganization energy, while slightly underestimating the SPME result in selected frequency ranges.

\subsection{Hole transport from FOB-SH simulations}
\label{sec:results_hole_transport}

To investigate how Coulombic interactions affect hole transport in rubrene, we performed FOB-SH simulations at $300$~K using the DSF electrostatics scheme. 
These results are compared directly with previously reported FOB-SH simulations for the same system using the zero-charge model\cite{elsner2024thermoelectric}.
Both sets of simulations used the same supercell size, \(50\times13\times1\) unit cells, corresponding to $1300$ electronically active molecules in the high-mobility \((a,b)\) plane.

At the microscopic level, charge transport is governed by a competition between terms in the electronic Hamiltonian that promote delocalization and those that induce dynamic localization. The off-diagonal matrix elements, ($H_{kl}, k\neq l$), describe electronic couplings between neighbouring molecules and drive spatial delocalization of the charge carrier. 
By contrast, thermal fluctuations of the diagonal site energies, ($H_{kk}$), and off-diagonal couplings constitute dynamic energetic disorder, which tends to localize the carrier\cite{fratini2016transient}.
The site energy fluctuations are directly related to the vertical energy-gap fluctuations and hence to the fluctuation-derived reorganization energy, Eq.~\ref{eq:lambda_var}, which was shown in Section~\ref{sec:results_reorganization_energy}
to moderately increase by
Coulombic interactions. We therefore first examine how the inclusion of electrostatics modifies the distributions of electronic couplings and site energies before analyzing the resulting charge carrier delocalization and mobility.

\begin{table*}[tbp]
  \centering
  \caption{Thermal averages and root-mean-square fluctuations of electronic couplings and site energies,
  average and root-mean-square fluctuations of the IPR, timescales associated with transient delocalization events, and hole mobility obtained from FOB-SH simulations at $300$~K with and without electrostatics.}
  \label{table:FOBSH_data}
  \begin{tabular}{lccccccccccc}
    \hline
     & $\langle J_a \rangle^a$ &  $\sigma_a^a$ & $\langle J_b \rangle^a$ & $\sigma_b^a$ & $\sigma_E^a$ & $\langle \mathrm{IPR} \rangle^b$ & $\sigma_\mathrm{IPR}^b$ & $t_r^c$ & $\tau_r^c$ & $\mu_a^d$ & $\mu_b^d$\\
    \hline
    Zero-charge model$^e$ & 106.8  &  28.2 & -15.6 & 6.7 & 63.5 & 13.0 & 10.0 & 9 & 71 &  34.8 $\pm$ 1.5 & 6.5 $\pm$ 0.4\\
    DSF &  103.4 &  27.8 &  -16.4  &  6.7  & 69.2 &  8.9 & 8.0 & 10 & 89 & 21.1 $\pm$ 2.0 & 5.0 $\pm$ 0.2\\
    Experiment$^f$ &   &   &    &   &  &  &  & & & 20.2 & 7.7\\
    \hline
  \end{tabular}
  \begin{flushleft}
    \small
    $^a$ Mean value of electronic couplings, $\langle J_{a(b)} \rangle$, and root-mean-square fluctuations of electronic couplings, $\sigma_{a(b)}$, and site energies, $\sigma_{E}$, in meV. \\
    $^b$ Average inverse participation ratio of the hole wavefunction, $\langle$IPR$\rangle$ (Eq.~\ref{eq:IPR_WFN}) and corresponding root-mean-square fluctuation over all trajectories, $\sigma_\mathrm{IPR}$, neglecting the first 500~fs of dynamics. \\
    $^c$ Average duration of a transient delocalization event, $t_r$ (fs), defined as time intervals during which the IPR of a trajectory is greater than $\langle$IPR$\rangle + \sigma_\mathrm{IPR}$, and inverse rate of transient delocalization events, $\tau_r$ (fs). \\
    $^d$ Charge carrier mobility from FOB-SH, Eq.~\ref{eq:Einstein_relation}, along the $a$ and $b$ crystallographic directions, in cm$^2$/Vs. \\
    $^e$ FOB-SH simulations without electrostatics, from Ref.~\citenum{elsner2024thermoelectric}.\\
    $^f$ Experimental results from Ref.~\citenum{podzorov2004intrinsic}.\\
  \end{flushleft}
\end{table*}

Table~\ref{table:FOBSH_data} summarizes the distributions of the electronic Hamiltonian matrix elements obtained from the FOB-SH simulations, including the mean electronic couplings along the \(a\) and \(b\) crystallographic directions, \(\langle J_{d}\rangle\) \((d=a,b)\), their root-mean-square fluctuations, \(\sigma_{d}\), and the root-mean-square fluctuations of the site energies, \(\sigma_E\). The table also reports the average inverse participation ratio (IPR), \(\langle\mathrm{IPR}\rangle\), which measures the spatial extent of the hole wavefunction, together with \(\sigma_{\mathrm{IPR}}\), the root-mean-square fluctuation of the IPR over the ensemble of trajectories. Finally, the table reports hole mobilities along the \(a\) and \(b\) directions, \(\mu_a\) and \(\mu_b\), obtained from the linear regime of the time-dependent mean-square displacement (MSD) using the Einstein-Smoluchowski relation (Eqs.~\ref{eq:MSD} and~\ref{eq:Einstein_relation}).

The mean electronic couplings are only weakly affected by the inclusion of DSF electrostatics, remaining within \(5\%\) of the zero-charge values along both crystallographic directions. Similarly, the root-mean-square fluctuations of the electronic couplings, \(\sigma_a\) and \(\sigma_b\), change by less than \(2\%\). 
The dominant effect of Coulombic interactions is instead an increase in the site energy disorder: \(\sigma_E\) increases by \(9\%\), from \(63.5\)~meV in the zero-charge model to \(69.2\)~meV with DSF electrostatics.

This increase in site energy disorder can be rationalized directly from the reorganization energies reported in Section~\ref{sec:results_reorganization_energy}. 
If the thermal fluctuations of adjacent molecular site energies, \(E_1\) and \(E_2\), are assumed to be statistically equivalent and approximately uncorrelated, \(\mathrm{Cov}(E_1,E_2)\approx 0\), then the variance of the vertical energy gap is
\begin{equation}
    \sigma_{\Delta E}^2
    =
    \mathrm{Var}(E_2-E_1)
    \approx
    2\sigma_E^2 .
\end{equation}
Combining this relation with Eq.~\ref{eq:lambda_var} gives
\begin{equation}
\label{eq:site_energy_lambda_var}
    \sigma_E
    \approx
    \sqrt{k_{\mathrm{B}}T\lambda^{\mathrm{var}}}.
\end{equation}
Using the values in Table~\ref{tab:lambda_reorganization_summary}, \(\lambda^{\mathrm{var}}=155\)~meV for the zero-charge model and an orientation-averaged value of \(\lambda^{\mathrm{var}}=180.5\)~meV for DSF, this expression predicts \(\sigma_E=63.3\)~meV and \(68.3\)~meV, respectively. These estimates agree closely with the site energy fluctuations sampled directly in the FOB-SH simulations, \(63.5\)~meV and \(69.2\)~meV. We note that this comparison is not fully consistent because the relation Eq.~\ref{eq:site_energy_lambda_var} assumes that both $\sigma_E$ and $\lambda^\mathrm{var}$ are obtained from ground state MD simulation in a charge localized state, whereas the $\sigma_E$ value reported here are obtained from FOB-SH non-adiabatic dynamics. Evidently this difference does not play a significant role here.

This agreement shows that the increase in site energy disorder observed in the transport simulations is quantitatively consistent with the electrostatic increase in the reorganization energy. Thus, the inclusion of Coulombic interactions primarily modifies the diagonal energetic disorder experienced by the hole, while leaving the distributions of electronic couplings largely unchanged.

\begin{figure*}[t]
\centering
\includegraphics[width=0.75\textwidth]{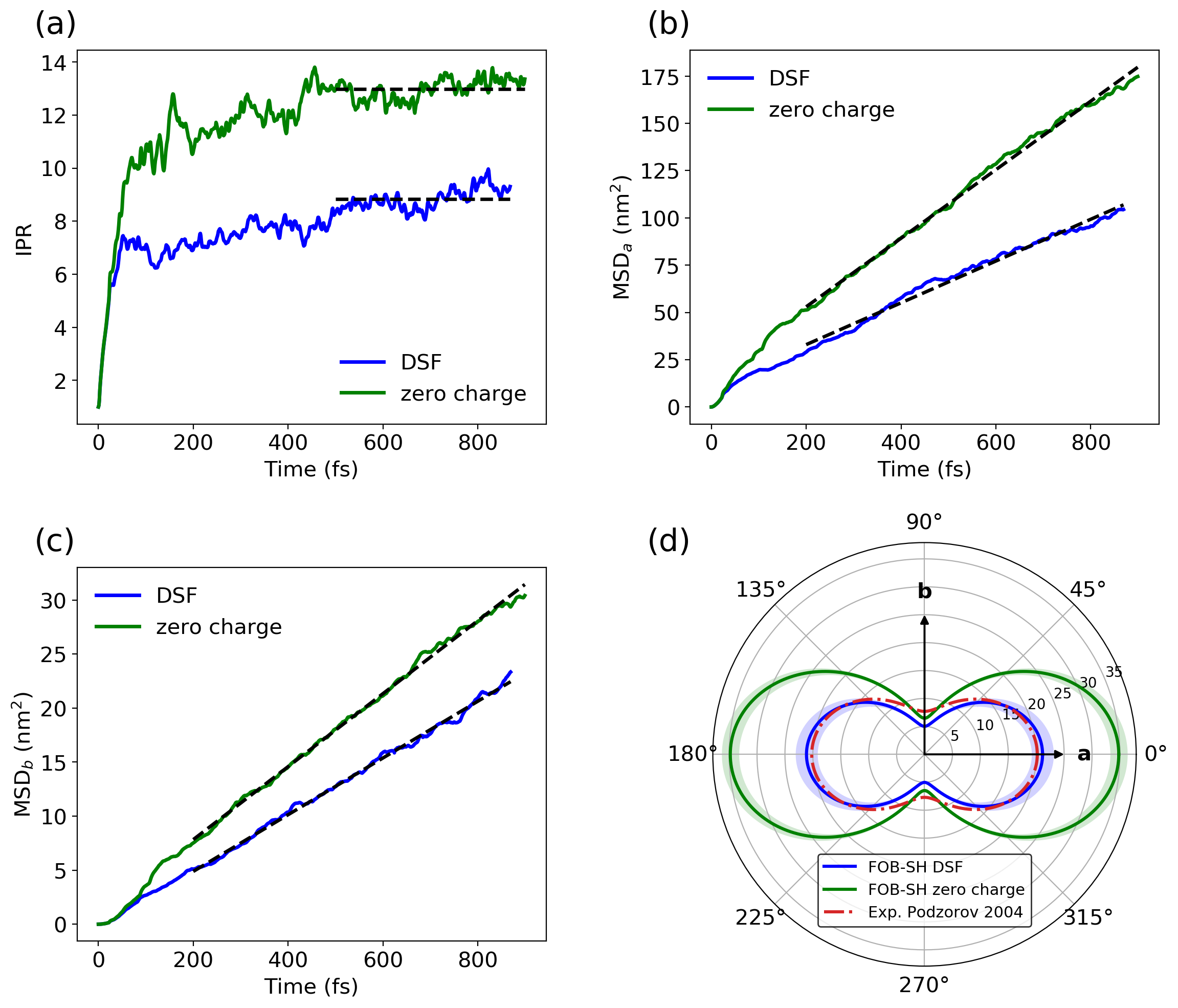}
  \caption{
  Overview of results from FOB-SH simulations of hole transport in rubrene at 300~K employing DSF electrostatics (blue) alongside results from Ref.~\citenum{elsner2024thermoelectric} using the zero-charge model (green). 
  (a) Time evolution of the average IPR, Eq.~\ref{eq:IPR_WFN}. Horizontal black dashed lines indicate the average value of the IPR over the interval $500-900$~fs. 
  (b, c) Mean-square displacement of the charge carrier wavefunction, Eq.~\ref{eq:MSD}, as a function of time along the $a$ and $b$ crystallographic directions, respectively. Linear fits used to calculate the charge carrier mobility are shown in black dashed lines. 
  (d) Polar representation of the charge carrier mobility with experimental values from Ref.~\citenum{podzorov2004intrinsic} in the red dash-dotted line. The $a$ and $b$ crystallographic directions correspond to $0^{\circ}$ and $90^{\circ}$, respectively.
}
  \label{fgr:mobility_MSD_IPR}
\end{figure*}

The enhanced site energy disorder has a direct impact on the spatial extent of the charge carrier wavefunction. Figure~\ref{fgr:mobility_MSD_IPR}(a) shows the time evolution of the IPR (Eq.~\ref{eq:IPR_WFN}), which measures the number of molecular sites over which the hole wavefunction is delocalized. In both the zero-charge and DSF simulations, the IPR, averaged over all FOB-SH trajectories, increases rapidly during the first \(\approx 200\)~fs, reflecting quantum relaxation from the initially localized state (see Section~\ref{sec:methods_fobsh_details}). At later times, the wavefunction remains substantially more delocalized in the zero-charge model, with an average IPR of $13$, compared to $9$ when DSF electrostatics are included.

This reduction in charge carrier delocalization is reflected in the MSD of the hole wavefunction. Figures~\ref{fgr:mobility_MSD_IPR}(b) and (c) show the MSD, averaged over all FOB-SH trajectories, along the \(a\) and \(b\) crystallographic directions, respectively. In both directions, the MSD grows more slowly when electrostatic interactions are included, indicating reduced charge diffusion. Consequently, the mobilities obtained from Eq.~\ref{eq:Einstein_relation} are lower for the DSF simulations than for the zero-charge model.

Figure~\ref{fgr:mobility_MSD_IPR}(d) shows the calculated mobility tensor in polar representation,
\(\mu(\theta)=\mu_a\cos^2\theta+\mu_b\sin^2\theta\), for the zero-charge and DSF simulations, together with the experimental values reported by Podzorov \emph{et al.}\cite{podzorov2004intrinsic}.
Without Coulombic interactions, the calculated mobility along the high-mobility \(a\) direction is \(34.8\)~cm\(^2\)/Vs\cite{elsner2024thermoelectric}, overestimating the experimental value of \(20.2\)~cm\(^2\)/Vs\cite{podzorov2004intrinsic} by a factor of \(\approx 1.7\). Including DSF electrostatics reduces this value to \(21.1\)~cm\(^2\)/Vs, bringing the simulation into near-quantitative agreement with experiment. 

Along the low mobility \(b\) direction, both models slightly underestimate the experimental mobility. However, including electrostatics improves the overall anisotropy of the mobility tensor. The anisotropy ratio decreases from \(\mu_a/\mu_b=5.4\) in the zero-charge model to \(4.2\) with DSF electrostatics, which is closer to the experimental ratio of \(2.6\)\cite{podzorov2004intrinsic}. 

The inclusion of electrostatic interactions does not qualitatively change the underlying transient delocalization mechanism. To quantify this, we identify transient delocalization events as time intervals during which the IPR of a given trajectory exceeds \(\langle\mathrm{IPR}\rangle+\sigma_{\mathrm{IPR}}\). The values of \(\langle\mathrm{IPR}\rangle\) and \(\sigma_{\mathrm{IPR}}\) are listed in Table~\ref{table:FOBSH_data}, along with the average duration of transient delocalization events, \(t_r\), and their inverse event rate, \(\tau_r\), i.e., a measure of the average time between two transient delocalization events. Both \(t_r\) and \(\tau_r\) are very similar in the zero-charge and DSF simulations: \(t_r\) increases only slightly from 9 to 10~fs, while \(\tau_r\) increases from 71 to 89~fs. Thus, while electrostatic interactions reduce the average spatial extent of the hole wavefunction by increasing site energy disorder, the transient delocalization mechanism is preserved.

\section{Conclusions}

In this work, we have investigated the role of Coulombic electrostatic interactions in determining the reorganization energy and hole mobility of rubrene at \(300\)~K. 
Including electrostatics leads to a moderate increase in
the reorganization energy relative to the zero-charge model, reflecting additional electrostatic contributions to vertical energy-gap fluctuations. 
Spectral analysis of these fluctuations shows that this increase is not confined to a single frequency range, but arises from enhanced coupling across both low-frequency nuclear motions and higher-frequency intramolecular vibrations. 
Benchmarking against SPME calculations further shows that the computationally efficient real-space DSF scheme captures the main electrostatic contribution to the reorganization energy, while slightly underestimating the SPME reference.

The increase in reorganization energy translates directly into enhanced site energy fluctuations in explicit FOB-SH simulations of hole transport. In contrast, the mean electronic couplings and their thermal fluctuations are only weakly affected by the inclusion of DSF electrostatics. The dominant effect of Coulombic interactions is therefore to increase diagonal energetic disorder, which reduces the spatial extent of the hole wavefunction. This is reflected in a decrease of the average IPR from \(13\) in the zero-charge model to \(9\) with DSF electrostatics.

This enhanced localization has a substantial effect on the predicted charge carrier mobility. Previous FOB-SH simulations using the zero-charge model yielded a hole mobility of \(34.8~\mathrm{cm^2\,V^{-1}\,s^{-1}}\) along the high-mobility \(a\) direction, overestimating the experimental reference by a factor of approximately \(1.7\). Including electrostatic site energy disorder through the DSF scheme reduces this value to \(21.1~\mathrm{cm^2\,V^{-1}\,s^{-1}}\), in close agreement with the experimental value of \(\approx 20~\mathrm{cm^2\,V^{-1}\,s^{-1}}\). The inclusion of electrostatics also improves the predicted mobility anisotropy. Overall, these results show that electrostatic site energy disorder is essential for quantitatively predictive simulations of hole transport in rubrene.

\section{Methods}

\subsection{Electrostatic Interactions}
\label{sec:electrostatic_interactions}

Electrostatic interactions were treated using the DSF method proposed by Fennell and Gezelter~\cite{fennell2006ewald}, which provides a computationally efficient alternative to conventional Ewald summation for periodic systems. In the DSF approach, the long-range Coulomb interaction is represented by a damped real-space interaction that is smoothly shifted to zero at a finite cutoff radius ($R_{\mathrm{cut}}$). The electrostatic energy is written as

\begin{equation}
E^{\text{DSF}}  =   \frac{1}{2} \sum_{i\!=\!1}^{N_{\text{at}}} \sum_{j\!=\!1}^{N_{\text{at}}} \sum_{\bf n}^\prime 
                                q_i q_j   R^{\text{DSF}} (| {\bf R}_{ij} + {\bf n} |  )  \label{dsf-energy}     
\end{equation}

where \(q_i\) and \(q_j\) are the atomic charges, \(\mathbf{R}_{ij}\) is the distance vector between atoms \(i\) and \(j\), and \(\mathbf{n}\) denotes lattice translation vectors under periodic boundary conditions, with the $^\prime$ symbol indicating that $i=j$ is excluded when ${\bf n} = {\bf 0}$. $R^{\text{DSF}}$ is the damped shifted interaction kernel including force shifting and real-space truncation, whose full expression is given in Ref.~\citenum{giannini2025efficient}.

Compared to Ewald summation, the DSF method avoids reciprocal-space calculations entirely and enables efficient evaluation of electrostatic site energies for multiple charge-localized states, leading to near-linear scaling with system size while retaining good accuracy for condensed-phase systems~\cite{fennell2006ewald,giannini2025efficient}. In particular, the DSF formalism allows us to use an efficient  addition--subtraction scheme (introduced in Ref.~\citenum{giannini2025efficient}) through which the electrostatic energy of a given charge-localized state is decomposed into the electrostatic energy of the fully neutral system plus correction terms accounting for the molecule carrying the excess charge. This decomposition reduces the computational overhead associated with calculating all $N_{\mathrm{mol}}$ charge states from $O(N_{\mathrm{mol}})$ to approximately $O(N_{\mathrm{mol}}^0)$, i.e., to a near system-size independent prefactor relative to the cost of a single electrostatic calculation.

Benchmark calculations performed in Ref.~\citenum{giannini2025efficient} demonstrated that DSF reproduces Ewald electrostatic energies and forces with good accuracy when suitable damping parameters and cutoff radii are employed. Using $R_{\mathrm{cut}} = 16$~\AA{} and a damping parameter $\alpha = 0.09$~\AA$^{-1}$, mean unsigned errors in electrostatic energies and forces remained below approximately $5\%$ relative to Ewald summation.

\subsection{Implicit Polarizable Force Field via Charge Scaling}
\label{sec:Implicit Polarizable Force Field via Charge Scaling}

In conventional fixed-charge force fields, the absence of electronic polarization is known to lead to an overestimation of electrostatic interactions, resulting in exaggerated site energy fluctuations and reorganization energies~\cite{warshel1982dynamics,gupta2004effects}. In particular, the outer-sphere contribution to the reorganization energy is typically overestimated. These effects are relevant in nonadiabatic molecular dynamics simulations of charge transport, where electrostatic disorder directly influences charge delocalization and mobility.

To account for electronic polarization effects in FOB-SH simulations while retaining the computational efficiency of a nonpolarizable force field, an implicit polarization scheme based on charge scaling was employed, following the strategy introduced in Ref.~\citenum{giannini2025efficient}. 
The atomic charges of the neutral molecular state, obtained from a DFT calculation of a rubrene molecule in vacuum, were uniformly scaled by a factor $\gamma$ according to
$q_i^{\mathrm{n,FF}} \rightarrow \tilde{q}_i^{\mathrm{n,FF}} = \gamma q_i^{\mathrm{n,FF}}$,
while the charges of the charged molecular state were defined as
$q_i^{\mathrm{c,FF}} \rightarrow \tilde{q}_i^{\mathrm{c,FF}} = \gamma q_i^{\mathrm{n,FF}} + \Delta q_i^{\mathrm{DFT}}$,
where $\Delta q_i^{\mathrm{DFT}} = q_i^{\mathrm{c,DFT}} - q_i^{\mathrm{n,DFT}}$ is the charge difference obtained from DFT calculations between charged and neutral states of a rubrene molecule in vacuum. Atomic charges for rubrene were obtained by fitting the electrostatic potential (ESP) using the Merz–Kollman scheme\cite{singh1984approach}. The ESP was computed for the optimized geometries of neutral and cationic rubrene in vacuum. All calculations were performed at the B3LYP/6-311G(d) level of theory.

Physically, charge scaling mimics dielectric screening arising from the fast electronic polarization of the surrounding medium. Reducing the magnitude of the electrostatic interactions decreases both the amplitude of site energy fluctuations and the external contribution to the reorganization energy.

The optimal value of $\gamma$ can be determined by matching the total reorganization energy obtained from scaled-charge simulations to the corresponding value computed using an explicitly polarizable force field with induced dipoles, as done in Ref.~\cite{giannini2025efficient}. For apolar systems such as anthracene, a value of $\gamma = 0.80$ was found to reproduce accurately the reorganization energies and electrostatic fluctuations obtained from polarizable molecular dynamics simulations, while maintaining a significantly lower computational cost. The same value of $\gamma$ was adopted here for rubrene.

\subsection{Reorganization energy}
\label{sec:methods_reorganization_energy}

The reorganization energy for hole transfer from molecule \(M\) to a neighbouring molecule \(N\) was computed from the distribution of vertical energy gaps evaluated along a classical molecular dynamics trajectory in which the hole was localized on molecule \(M\). The vertical energy gap is defined as
\begin{equation}
\label{eq:vertical_energy_gap}
    \Delta E_{M \rightarrow N}(t)
    =
    E_N(\mathbf{R}(t)) - E_M(\mathbf{R}(t)),
\end{equation}
where \(\mathbf{R}(t)\) denotes a nuclear configuration sampled from the trajectory in which the hole was localized on molecule \(M\). Here, \(E_M(\mathbf{R}(t))\) is the potential energy of the system with the hole localized on molecule \(M\), while \(E_N(\mathbf{R}(t))\) is the potential energy of the same nuclear configuration with the hole localized on molecule \(N\). 

The Stokes-shift reorganization energy is given by\cite{blumberger2015recent}
\begin{equation}
\label{eq:lambda_st}
    \lambda^{\mathrm{st}}
    =
    \left\langle
    \Delta E_{M\rightarrow N}
    \right\rangle_M ,
\end{equation}
where \(\langle \cdots \rangle_M\) denotes an average over nuclear configurations sampled on the diabatic state with the hole localized on molecule~\(M\). 
The reorganization energy can also be computed based on the fluctuations of the vertical energy gap\cite{blumberger2015recent}, 
\begin{equation}
\label{eq:lambda_var}
    \lambda^{\mathrm{var}}
    =
    \frac{
    \left\langle
    \left[
    \delta \Delta E_{M\rightarrow N}
    \right]^2 \right\rangle_M
    }{2 k_{\mathrm{B}} T},
\end{equation}

with 
\begin{equation}
    \delta \Delta E_{M\rightarrow N}(t)
    =
    \Delta E_{M\rightarrow N}(t)
    -
    \left\langle
    \Delta E_{M\rightarrow N}
    \right\rangle_M.
\end{equation}

In the linear response regime, \(\lambda^{\mathrm{st}}\) and \(\lambda^{\mathrm{var}}\) are identical.

For each treatment of electrostatics, the system was first equilibrated to \(300\)~K on the diabatic state where molecule \(M\) is in the \(+1\) charge state for 500 ps in the NVT ensemble using a Nosé--Hoover thermostat. Production trajectories were then generated on this diabatic state for 1 ns in the NVE ensemble, with configurations saved every 4 fs. 
\(E_N(\mathbf{R}(t))\) was computed by single-point energy recalculations over the saved trajectory by transferring the cationic parameters from molecule \(M\) to molecule \(N\) for the molecules \(N=P1, P2, T1, T2\) depicted in Figure~\ref{fgr:crystal_P_T}.
In practice, localization of the hole on a given molecule was represented by assigning that molecule force-field parameters corresponding to the \(+1\) charge state, with all other molecules described by parameters corresponding to the neutral state. 

For each transfer direction, the reported reorganization energy was obtained by computing \(\lambda^{j}\), \(j=\mathrm{st,var}\), separately for the two symmetry-related dimers and averaging the two values. The quoted uncertainty combines the block-bootstrap sampling uncertainty within each trajectory with the residual scatter between the two symmetry-related dimers. Further details of the uncertainty estimate, together with the individual dimer values, are provided in the Supplementary Section~1.

To analyze the frequency dependence of the energy gap fluctuations, we computed the autocorrelation function
\begin{equation}
    C(t)
    =
    \left\langle
    \delta \Delta E_{M\rightarrow N}(0)
    \delta \Delta E_{M\rightarrow N}(t)
    \right\rangle_M.
\end{equation}

The spectral density of vertical energy gap fluctuations is given by
\begin{equation}
\label{eq:spectral_density}
    \frac{S(\omega)}{\omega}
    =
    \frac{\beta}{4}
    \int_0^\infty
    C(t)\cos(\omega t)\,dt,
\end{equation}
where \(\beta=(k_{\mathrm{B}}T)^{-1}\). The cumulative fluctuation-derived reorganization energy was then evaluated as
\begin{equation}
\label{eq:lambda_cumulative}
    \lambda^{\mathrm{cum}}(\omega)
    =
    \frac{4}{\pi}
    \int_0^\omega
    \frac{S(\omega')}{\omega'}\,d\omega'.
\end{equation}

By construction, integration over all frequencies recovers the fluctuation estimate,
\begin{equation}
\lambda^{\mathrm{var}}
    =
    \lim_{\omega\to \infty}
    \lambda^{\mathrm{cum}}(\omega).
\end{equation}

For the spectra shown in Figure~\ref{fgr:spectral_density_distributions}(a,b), 
autocorrelation functions were computed from 15~ps sliding windows advanced in 4~fs increments and averaged over all windows. The resulting averaged autocorrelation function was tapered with a Hann window to reduce finite-window artefacts in \(S(\omega)/\omega\). For the cumulative reorganization energies shown in Figure~\ref{fgr:spectral_density_distributions}(c,d), the autocorrelation function was computed over the full trajectory and cosine transformed without applying a Hann taper, so that the integrated spectral weight is consistent with \(\lambda^{\mathrm{var}}\).

\subsection{Fragment orbital-based surface hopping}

Charge transport simulations were performed using FOB-SH, a mixed quantum--classical dynamics method based on fewest-switches surface hopping\cite{spencer2016fob,carof2019calculate}. In this approach, a single excess charge carrier is propagated quantum mechanically in a basis of orthogonalized, time-dependent frontier orbitals associated with the molecules of the crystal. For hole transport, the wavefunction is expanded in the orthogonalized HOMOs of the molecular fragments,
\begin{equation}
    \ket{\Psi(t)}
    =
    \sum_{l=1}^{M}
    u_l(t)
    \ket{\phi_l(\mathbf{R}(t))},
    \label{eq:wavefunction}
\end{equation}
where \(\mathbf{R}(t)\) denotes the nuclear configuration.

The charge carrier evolves under the following time-dependent Hamiltonian,
\begin{equation}
    H(t)
    =
    \sum_k
    \epsilon_k(t)
    \ket{\phi_k(t)}\bra{\phi_k(t)}
    +
    \sum_{k\neq l}
    H_{kl}(t)
    \ket{\phi_k(t)}\bra{\phi_l(t)},
    \label{eq:Hamiltonian}
\end{equation}
where \(\epsilon_k\) are site energies and \(H_{kl}\) are electronic couplings. Site energies are evaluated using a classical force field by assigning the charged-state parameters to molecule \(k\), with all other molecules in the neutral state. Electronic couplings were computed using the analytic overlap method (AOM)\cite{ziogos2021ultrafast}, in which the electronic coupling is assumed to scale linearly with the overlap between fragment orbitals projected onto a minimal Slater-type orbital basis. The proportionality constant was obtained by fitting to reference couplings computed using projector-operator-based diabatization\cite{futera2017electronic}.

The wavefunction coefficients \(u_l(t)\) are propagated according to the time-dependent Schr{\"o}dinger equation in the (quasi-diabatic) site basis,
\begin{equation}
    i\hbar \dot{u}_k(t)
    =
    \sum_{l=1}^{M}
    u_l(t)
    \left[
    H_{kl}(\mathbf{R}(t))
    -
    i\hbar d_{kl}(\mathbf{R}(t))
    \right],
    \label{eq:TDSE}
\end{equation}
where \(d_{kl}=\langle \phi_k | \dot{\phi}_l\rangle\) are the nonadiabatic coupling elements in the site basis. At each nuclear time step, nuclei are propagated classically on a single active adiabatic surface. Stochastic hops between adiabatic states occur according to the fewest-switches surface-hopping criterion\cite{tully1990molecular}. 
After a surface hop, energy conservation is enforced by rescaling the nuclear velocities along the direction of the nonadiabatic coupling vector (NACV). If the available nuclear kinetic energy is insufficient to conserve the total energy, the hop is rejected and the velocity component along the NACV is reversed\cite{carof2019calculate,carof2017detailed}.

Several extensions to the standard surface-hopping algorithm were employed, including decoherence correction, trivial-crossing detection, and suppression of decoherence-induced spurious long-range charge transfer. These corrections are required to obtain converged mobilities, detailed balance, and internally consistent charge dynamics; detailed descriptions are given in Refs.~\citenum{carof2019calculate} and \citenum{carof2017detailed}. 

\subsection{FOB-SH Simulation details}
\label{sec:methods_fobsh_details}

Interatomic interactions in rubrene were described using a force field based on the general amber force field (GAFF)\cite{wang2004development}, with selected parameters re-optimized to better reproduce the thermal distribution of electronic couplings obtained from ab-initio MD simulations\cite{elsner2021mechanoelectric}, see Ref.~\citenum{elsner2024thermoelectric} for details.
Electrostatic interactions were treated as explained in Sections~\ref{sec:electrostatic_interactions} and ~\ref{sec:Implicit Polarizable Force Field via Charge Scaling}.
A supercell of $50\times13\times1$ rubrene unit cells was equilibrated to 300~K for 200~ps using a time step of 1~fs in the NVT ensemble using a Nos\'e-Hoover thermostat. 
Initial positions and velocities for the swarm of FOB-SH trajectories were drawn from snapshots separated by 0.5 ps from a subsequent 250~ps MD trajectory in the NVE ensemble.
A total of 460 FOB-SH trajectories were propagated. 
For each FOB-SH trajectory, the wavefunction was initialised on a single molecular site (i.e. diabatic state) located at the corner of the simulation box and propagated for 900 fs in the NVE ensemble using an MD time step of 0.05 fs and an electronic time step of 0.01 fs for integration of Eq.~\ref{eq:TDSE} employing the Runge-Kutta algorithm to 4th order. 
The initial 200 fs of dynamics, corresponding to quantum relaxation from the initial diabatic state to the partially delocalized polaronic state, was neglected in all analysis. 
Average IPR values and transient delocalization statistics were computed over the interval \(500\text{--}900\)~fs.
Results for the model without Coulombic electrostatics, denoted the ``zero-charge model'', are reproduced from Ref.~\citenum{elsner2024thermoelectric}. In that work, the same supercell size, electronically active region, initialization procedure, and analysis protocol were used, with averages taken over 686 FOB-SH trajectories.

\subsection{Calculation of charge mobility and IPR}

Hole mobilities are obtained from FOB-SH trajectories from the mean-square displacement (MSD) of the hole wavefunction, given by
\begin{align}
\mathrm{MSD}_{\alpha\beta}(t)
&=
\frac{1}{N_{\mathrm{traj}}}
\sum_{n=1}^{N_{\mathrm{traj}}}
\bra{\Psi_n(t)}
(\alpha-\alpha_{0,n})(\beta-\beta_{0,n})
\ket{\Psi_n(t)}
\\
&\approx
\frac{1}{N_{\mathrm{traj}}}
\sum_{n=1}^{N_{\mathrm{traj}}}
\sum_{k=1}^{M}
|u_{k,n}(t)|^2
\left[\alpha_{k,n}(t)-\alpha_{0,n}\right]
\left[\beta_{k,n}(t)-\beta_{0,n}\right],
\label{eq:MSD}
\end{align}
where \(\Psi_n(t)\) is the hole wavefunction in trajectory \(n\), \(N_{\mathrm{traj}}\) is the number of FOB-SH trajectories, and \(\alpha,\beta\) denote Cartesian coordinates along the crystallographic \(a\) and \(b\) directions. In the discretized expression, Eq.~\ref{eq:MSD}, the hole position is represented by the molecular centres of mass, with \(\alpha_{k,n}(t)\) the \(\alpha\)-coordinate of molecule \(k\) in trajectory \(n\), and \(|u_{k,n}(t)|^2\) the corresponding hole population. The quantity \(\alpha_{0,n}\) is the initial centre-of-charge position of the hole wavefunction along the \(\alpha\) direction in trajectory~\(n\), with \(\beta_{0,n}\) defined analogously.

The diffusion tensor is obtained from the long-time slope of the MSD,
\begin{equation}
D_{\alpha\beta}
=
\frac{1}{2}
\lim_{t\rightarrow\infty}
\frac{\mathrm{d}\mathrm{MSD}_{\alpha\beta}(t)}{\mathrm{d}t},
\label{eq:Diffusion_tensor}
\end{equation}
and the mobility tensor is then calculated using the Einstein-Smoluchowski relation,
\begin{equation}
\mu_{\alpha\beta}
=
\frac{eD_{\alpha\beta}}{k_{\mathrm{B}}T},
\label{eq:Einstein_relation}
\end{equation}
where \(e\) is the elementary charge, \(k_{\mathrm{B}}\) is the Boltzmann constant, and \(T\) is the temperature.

Delocalization of the time-dependent hole wavefunction is quantified using the inverse participation ratio (IPR),
\begin{equation}
\mathrm{IPR}(t)
=
\frac{1}{N_{\mathrm{traj}}}
\sum_{n=1}^{N_{\mathrm{traj}}}
\frac{1}{
\sum_{k=1}^{M}
|u_{k,n}(t)|^4
},
\label{eq:IPR_WFN}
\end{equation}
where \(M\) is the number of molecules in the electronically active region.

\section*{Author Contributions}
J.E.: Conceptualization, Data curation, Formal analysis, Investigation (lead), Methodology, Visualization, Writing – original draft, Writing – review \& editing. S.G.: Conceptualization, Investigation (supporting), Methodology, Software, Writing – review \& editing. J.B.: Conceptualization, Funding acquisition, Resources, Methodology, Supervision, Writing – review \& editing.

\section*{Conflicts of interest}

There are no conflicts to declare.

\section*{Data availability}

All data needed to evaluate the conclusions in this paper are present in the paper and/or the Supplementary Materials. 

\section*{Acknowledgements}

J. B. would like to acknowledge the UKRI for funding the ERC Advanced Grant ``EXCITING" under the Horizon Europe Guarantee (EP/Z533932/1).
Via our membership of the UK’s HEC Mate-
rials Chemistry Consortium, which is funded by EPSRC (Grant
Nos. EP/L000202 and EP/R029431), this work used the ARCHER2
UK National Supercomputing Service (http://www.archer2.ac.uk) as
well as the UK Materials and Molecular Modeling (MMM) Hub,
which is partially funded by EPSRC (Grant No. EP/P020194), for
computational resources.
S.G. acknowledges the Italian Ministry of University and Research for funding provided through the European Union-NextGenerationEU-PNRR program: Funding projects presented by young researchers Mission 4, Component 2, Investment Line 1.2.

\bibliographystyle{rsc}
\bibliography{rsc}

\beginsupplement

\begin{center}
{\Large \textbf{Supplementary Information}}\\[1em]

{\large \textbf{Impact of dynamic electrostatic disorder on hole mobility in rubrene: a nonadiabatic molecular dynamics investigation}}\\[1em]

Jan Elsner,$^{a,b,c}$ Samuele Giannini,$^{d}$ and Jochen Blumberger$^{a,*}$\\[0.75em]

{\small
$^{a}$Department of Physics and Astronomy and Thomas Young Centre, University College London, London WC1E 6BT, U.K.\\
$^{b}$Lehrstuhl f\"ur Theoretische Chemie II, Ruhr-Universit\"at Bochum, 44780 Bochum, Germany\\
$^{c}$Research Center Chemical Sciences and Sustainability, Research Alliance Ruhr, 44780 Bochum, Germany\\
$^{d}$Department of Chemistry and Industrial Chemistry, University of Pisa, 56124 Pisa, Italy\\
$^{*}$Corresponding author: j.blumberger@ucl.ac.uk
}
\end{center}

\vspace{1em}

\section{Reorganization energy for individual dimers}
\label{sec:suppl_reorg_indiv}

Table~\ref{tab:lambda_reorganization_summary} of the main text
(Section~\ref{sec:results_reorganization_energy}) reports reorganization
energies for hole transfer for $P$- and $T$-type dimers, where in each
case results are averaged over two symmetry-related dimers, shown in
Fig.~\ref{fgr:crystal_P_T} of the main text. Results for the individual dimers $P1$, $P2$,
$T1$, and $T2$, as well as the corresponding averaged values for $P$ and
$T$, are listed in Table~\ref{tab:lambda_reorganization_summary_indiv}.

\begin{table}[htbp!]
\centering
\caption{
Computed reorganization energies for hole transfer along individual dimers
$P1$, $P2$, $T1$, and $T2$ in rubrene using different treatments of
electrostatic interactions. The direction-averaged values $P$ and $T$ are
also shown.
}
\begin{tabular}{llcc}
\hline
Method & Dimer$^a$ &
$\lambda^{\mathrm{st}}$ / meV$^{b}$ &
$\lambda^{\mathrm{var}}$ / meV$^{c}$ \\
\hline
Zero charge & \(P1\) & $152$ & $156 \pm 3$ \\
Zero charge & \(P2\) & $152$ & $154 \pm 3$ \\
Zero charge & \(P\) & $152$ & $155 \pm 2$ \\
\hline
Zero charge & \(T1\) & $152$ & $160 \pm 4$ \\
Zero charge & \(T2\) & $152$ & $151 \pm 3$ \\
Zero charge & \(T\) & $152$ & $155 \pm 5$ \\
\hline
DSF & \(P1\) & $175$ & $176 \pm 3$ \\
DSF & \(P2\) & $175$ & $172 \pm 4$ \\
DSF & \(P\) & $175$ & $174 \pm 3$ \\
\hline
DSF & \(T1\) & $182$ & $190 \pm 4$ \\
DSF & \(T2\) & $182$ & $185 \pm 4$ \\
DSF & \(T\) & $182$ & $187 \pm 4$ \\
\hline
SPME & \(P1\) & $181$ & $190 \pm 4$ \\
SPME & \(P2\) & $181$ & $189 \pm 4$ \\
SPME & \(P\) & $181$ & $190 \pm 3$ \\
\hline
SPME & \(T1\) & $191$ & $197 \pm 4$ \\
SPME & \(T2\) & $191$ & $196 \pm 4$ \\
SPME & \(T\) & $191$ & $197 \pm 3$ \\
\hline
\end{tabular}
\begin{flushleft}
$^{a}$ Dimer considered for hole transfer (see Fig.~\ref{fgr:crystal_P_T} of the main text). $P$ and $T$ denote the averages over the two symmetry-related dimers, as reported in Table~\ref{tab:lambda_reorganization_summary} of the main text.
\\
$^{b}$ Stokes shift reorganization energy, $\lambda^\mathrm{st}$, Eq.~\ref{eq:lambda_st} of the main text. All statistical uncertainties for $\lambda^\mathrm{st}$ are smaller than 1~meV.\\
$^{c}$ Fluctuation-derived reorganization energy, $\lambda^\mathrm{var}$, Eq.~\ref{eq:lambda_var} of the main text, and statistical uncertainty. For individual dimer rows, uncertainties correspond to block-bootstrap standard errors. For direction-averaged \(P\) and \(T\) rows, uncertainties include both the finite-sampling contribution and the dimer-to-dimer scatter, as described below.
\end{flushleft}
\label{tab:lambda_reorganization_summary_indiv}
\end{table}

For each method and transfer direction, reorganization energies were first computed separately for the two symmetry-related dimers. 
The reported direction-averaged value was then calculated as
\begin{equation*}
    \bar{\lambda}^{j}
    =
    \frac{1}{2}
    \left(
    \lambda^{j}_{1}
    +
    \lambda^{j}_{2}
    \right),
    \qquad j\in\{\mathrm{st},\mathrm{var}\},
\end{equation*} 
where \(\lambda^{j}_{1}\) and \(\lambda^{j}_{2}\) denote the values for the two symmetry-related dimers for a given direction.

Uncertainties were estimated as follows. For each individual dimer,
finite-sampling uncertainties were obtained by block bootstrapping the
corresponding \(\Delta E_{M\rightarrow N}(t)\) trajectory. Each trajectory (total length 1~ns, saved every 4~fs)
was divided into non-overlapping contiguous 20~ps blocks. For each bootstrap replica,
blocks were resampled with replacement, using the same number of blocks as
in the original trajectory, and \(\lambda^{j}\) was recomputed. The standard
deviation of this bootstrap distribution gives the uncertainty reported for
the individual-dimer rows in Table~\ref{tab:lambda_reorganization_summary_indiv}, where shown.

For the direction-averaged \(P\) and \(T\) values, blocks were resampled with replacement independently for the two dimers, \(\lambda^{j}\) was recomputed for each resampled dimer trajectory, and the two resulting values were averaged. The standard deviation of this bootstrap distribution was taken as the finite-sampling standard error of the direction-averaged value, \(\sigma_{\mathrm{boot}}\).

Because the two dimers are symmetry-related but sampled by finite
trajectories, we also included the residual dimer-to-dimer scatter in the
uncertainty estimate. For two dimers, this contribution is
\[
\sigma_{\mathrm{dimer}}
=
\frac{
|\lambda^{j}_{1}-
\lambda^{j}_{2}|
}{2}.
\] 
The final standard error reported for the direction-averaged \(P\) and \(T\) values was obtained by adding the two contributions in quadrature,
\[
\sigma_{\mathrm{tot}}
=
\sqrt{
\sigma_{\mathrm{boot}}^{2}
+
\sigma_{\mathrm{dimer}}^{2}
}.
\]

We note that for $\lambda^\mathrm{st}$, which converges much faster than $\lambda^\mathrm{var}$, all statistical uncertainties are well below 1~meV and are therefore not reported.

\begin{figure*}[t]
\centering
\includegraphics[width=0.8\textwidth]{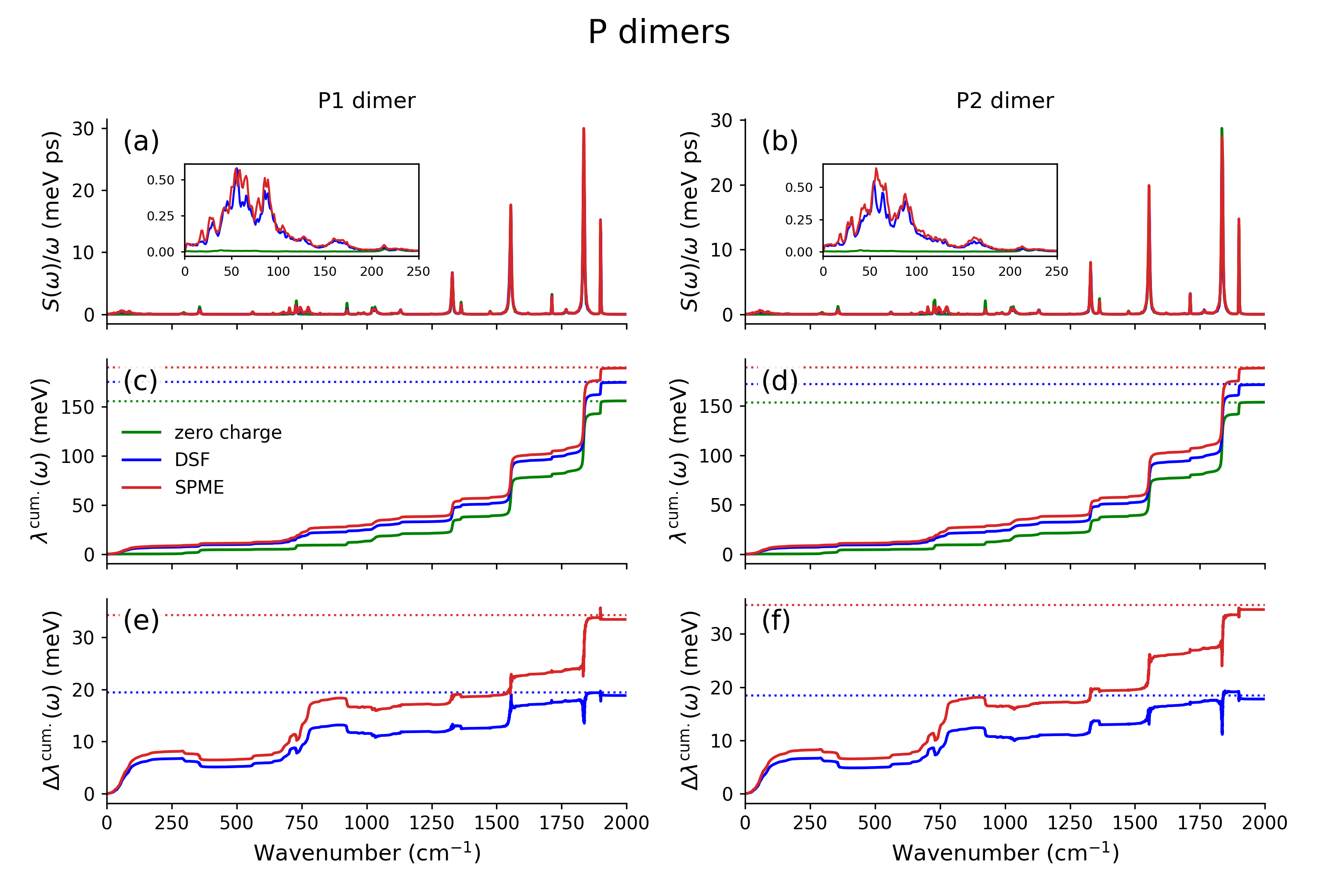}
  \caption{
Spectral density functions of vertical energy gap fluctuations and cumulative reorganization energies for hole transfer using different treatments of Coulombic interactions for $P$-type dimers.
(a,b) Spectral densities for transfer along the \(P1\) and \(P2\) directions, respectively. Insets highlight the low-frequency region from 0 to 250~cm\(^{-1}\).
(c,d) Cumulative fluctuation-derived reorganization energies up to frequency $\omega$, \(\lambda^\mathrm{cum}(\omega)\), for transfer along the \(P1\) and \(P2\) directions, respectively. The dotted horizontal lines indicate the corresponding $\lambda^\mathrm{var}$ values obtained from the full energy-gap distribution.
(e,f) Electrostatic contributions to the cumulative reorganization energy $\Delta\lambda^{\mathrm{cum}}(\omega) = \lambda^{\mathrm{cum}}_{j}(\omega) - \lambda^{\mathrm{cum}}_{\mathrm{zero\ charge}}(\omega)$ ($j = \text{DSF, SPME}$ in blue and red, respectively), for transfer along the \(P1\) and \(P2\) directions, respectively. Dotted lines indicate the values of $\Delta\lambda^{\mathrm{var}} = \lambda^{\mathrm{var}}_{j} - \lambda^{\mathrm{var}}_{\mathrm{zero\ charge}}$ ($j = \text{DSF, SPME}$ in blue and red, respectively).
}
  \label{fgr:spectral_density_P}
\end{figure*}

\begin{figure*}[t]
\centering
\includegraphics[width=0.8\textwidth]{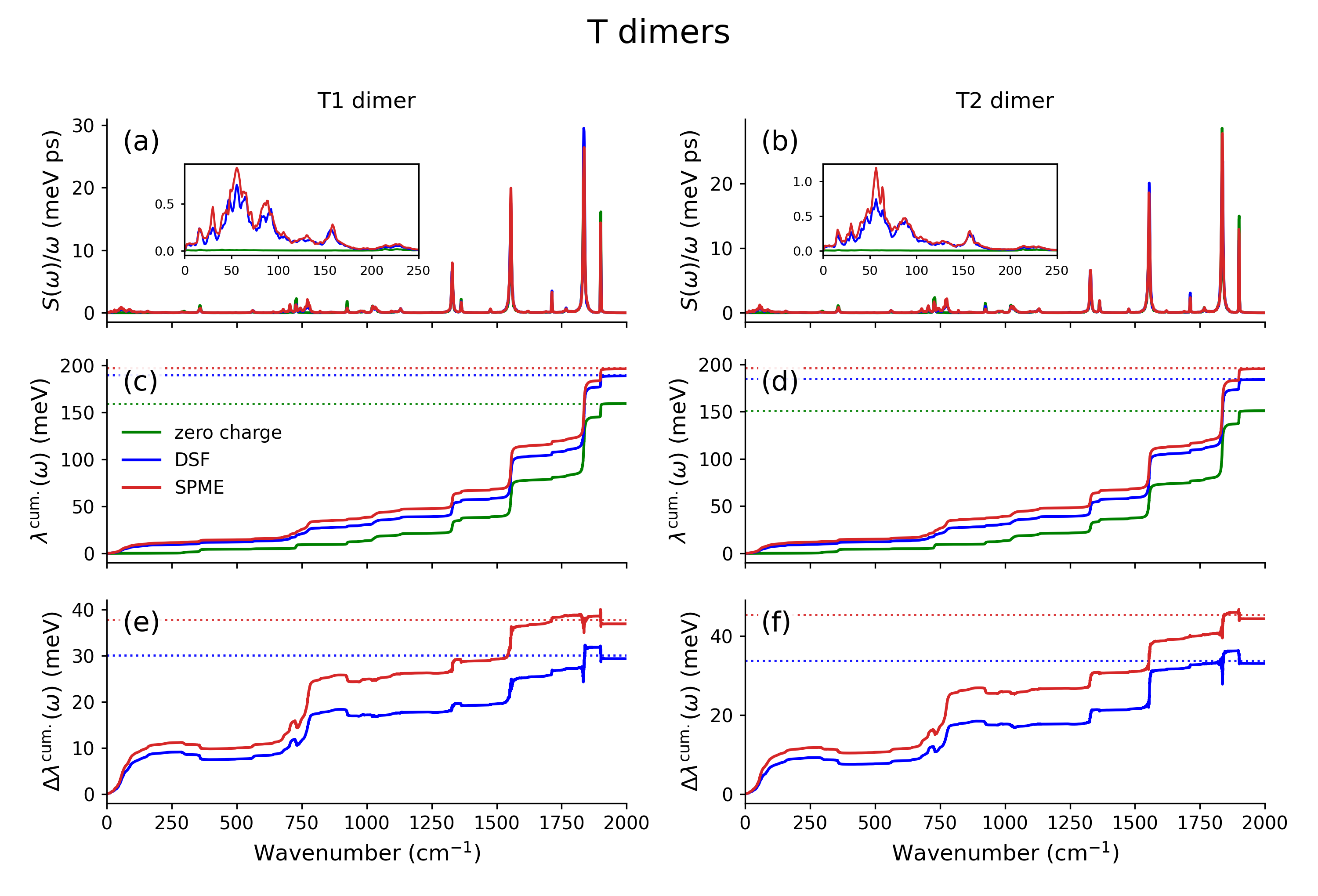}
  \caption{
Spectral density functions of vertical energy gap fluctuations and cumulative reorganization energies for hole transfer using different treatments of Coulombic interactions for $T$-type dimers.
(a,b) Spectral densities for transfer along the \(T1\) and \(T2\) directions, respectively. Insets highlight the low-frequency region from 0 to 250~cm\(^{-1}\).
(c,d) Cumulative fluctuation-derived reorganization energies up to frequency $\omega$, \(\lambda^\mathrm{cum}(\omega)\), for transfer along the \(T1\) and \(T2\) directions, respectively. The dotted horizontal lines indicate the corresponding $\lambda^\mathrm{var}$ values obtained from the full energy-gap distribution.
(e,f) Electrostatic contributions to the cumulative reorganization energy $\Delta\lambda^{\mathrm{cum}}(\omega) = \lambda^{\mathrm{cum}}_{j}(\omega) - \lambda^{\mathrm{cum}}_{\mathrm{zero\ charge}}(\omega)$ ($j = \text{DSF, SPME}$ in blue and red, respectively), for transfer along the \(T1\) and \(T2\) directions, respectively. Dotted lines indicate the values of $\Delta\lambda^{\mathrm{var}} = \lambda^{\mathrm{var}}_{j} - \lambda^{\mathrm{var}}_{\mathrm{zero\ charge}}$ ($j = \text{DSF, SPME}$ in blue and red, respectively).
}
  \label{fgr:spectral_density_T}
\end{figure*}

Figures~\ref{fgr:spectral_density_P} and \ref{fgr:spectral_density_T} show the spectral density functions of vertical energy gap fluctuations and the cumulative reorganization energies for the individual \(P\)- and \(T\)-type dimers, respectively. These plots correspond to Fig.~\ref{fgr:spectral_density_distributions} of the main text, but are resolved into the individual dimers \(P1/P2\) and \(T1/T2\). The two symmetry-related dimers show very similar spectral features and cumulative reorganization energies in each case.

\end{document}